# >3kV NiO/Ga$_2$O$_3$ Heterojunction Diodes with Space-Modulated Junction Termination Extension and Sub-1V Turn-on

Advait Gilankar, *Student Member, IEEE*, Abishek Katta, Nabasindhu Das, and Nidhin Kurian Kalarickal

*Abstract*—This work demonstrates high-performance vertical NiO/Ga$_2$O$_3$ heterojunction diodes (HJDs) with a 2-step space-modulated junction termination extension. Distinct from the current state-of-the-art Ga$_2$O$_3$ HJDs, we achieve breakdown voltage exceeding 3 kV with a low turn on voltage (V$_{ON}$) of 0.8V, estimated at a forward current density (I$_F$) of 1 A-cm$^{-2}$. The measured devices exhibit excellent turn-on characteristics achieving 100 A-cm$^{-2}$ current density at a forward bias of 1.5V along with a low differential specific on-resistance (R$_{on,sp}$) of 4.4 mΩ-cm$^2$. The SM-JTE was realized using concentric NiO rings with varying widths and spacing that approximates a gradual reduction in JTE charge. The unipolar figure of merit (FOM) calculated exceeds 2 GW-cm$^2$ and is among the best reported for devices with a sub-1V turn-on. The fabricated devices also displayed minimal change in forward I-V characteristics post reverse bias stress of 3 kV applied during breakdown voltage testing.

## I. Introduction

OVER the past decade, β-Ga$_2$O$_3$ has emerged as a prime candidate for next generation power electronics, owing to the high critical electric field (8 MV/cm), ease of n-type doping, and availability of large-area single crystal substrates [1], [2]. The availability of high quality, low defect density β-Ga$_2$O$_3$ drift-layers grown on bulk β-Ga$_2$O$_3$ substrates coupled with advanced damage-free etching techniques are paving the way for high-performance Ga$_2$O$_3$ power devices [3], [4], [5]. In particular, two-terminal vertical β-Ga$_2$O$_3$ devices have shown great improvements in performance over the past few years, exceeding the unipolar FOM of both SiC and GaN [6]. Majority of these reports employ p-NiO/Ga$_2$O$_3$ heterojunctions that have demonstrated superior breakdown performance [7], [8]. However, a key limitation with these HJDs is the much larger turn on voltage (V$_{ON}$) of ~2V. A higher V$_{ON}$ results in significantly higher conduction losses (~I$_{ON}^2$V$_{ON}$), degrading device efficiency. Therefore, achieving high breakdown voltage while maintaining low V$_{ON}$ is critical for enabling low-loss β-Ga$_2$O$_3$ rectifiers.

The work in ASU NanoFab was supported in part by the National Science Foundation Program No. NNCI-ECCS-1542160.
Advait Gilankar, Abishek Katta, Nabasindhu Das, and Nidhin Kurian Kalarickal are with the School of Electrical, Computer, and Energy Engineering, Arizona State University, Tempe, AZ 85287 USA (e-mail: agilanka@asu.edu).

Designing effective edge-termination structures with high junction termination efficiency is critical for achieving high reverse breakdown voltage in vertical devices. Several edge-termination structures including field-plates, ion-implantation, high-*k* dielectrics, junction termination extension (JTE) and mesa-etch termination have been reported in β-Ga$_2$O$_3$ diodes [9], [10], [11], [12], [13], [14], [15], [16], [17], [18], [19], [20], [21], [22], [23], [24], [25], [26]. Junction termination extensions (JTE) using p-type layers is an effective choice for obtaining high breakdown performance since it can achieve high JTE efficiency exceeding 90% [27]. JTEs, or advanced designs of JTEs are widely used in both Si and SiC vertical device technologies. In recent past, multi-zone and space-modulated JTE (SM-JTE) terminations have been used to demonstrate record breakdown performances in SiC diodes [28]. These structures are proven to provide wider fabrication windows and uniform distribution of electric field at the device edge. The SM-JTE structure results in similar JTE efficiency as that of a multi-zone JTE but can be realized using a smaller number of fabrication steps.

In this letter, we demonstrate vertical β-Ga$_2$O$_3$ HJDs employing the use of sputtered p-NiO to demonstrate breakdown voltage exceeding 3 kV. We leverage charge-balancing with SM-JTE design for achieving high blocking voltage. The fabricated diodes, showcase low forward V$_{ON}$, resulting in one of the lowest effective R$_{on}$ reported for NiO/Ga$_2$O$_3$ HJDs. The experimental results are supported by TCAD silvaco simulations, which indicate parallel plane junction field of ~3.8 MV/cm.

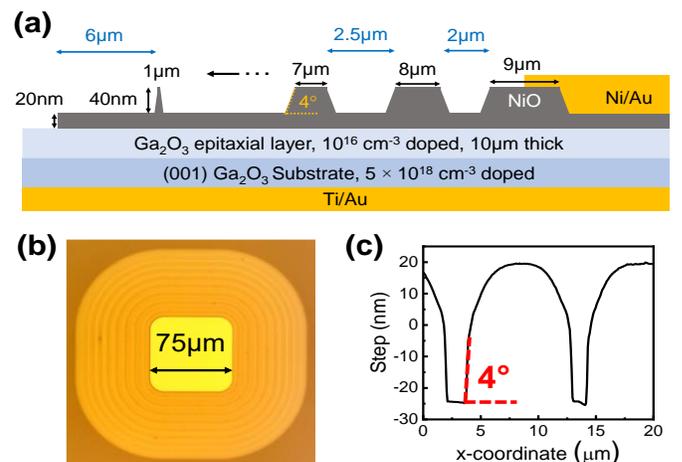

Fig. 1. (a) Cross-section schematic of fabricated p-NiO/Ga$_2$O$_3$ diode with SM-JTE termination. (b) Top-view optical image of fabricated 75µm active area device. (c) AFM Scan across the JTE structure showing a bevel angle of ~4° for NiO JTE rings.

## II. SM-JTE Design and Device Fabrication

The cross-sectional schematic of the fabricated device is shown in Fig. 1(a). The fabricated SM-JTE structure consists of nine p-NiO rings with gradually decreasing width and gradually increasing spacing between the rings. The width of the JTE rings increased from 9 μm near the anode to 1 μm at the edge of the device. Similarly, the spacing between the rings gradually increased from 1.5 μm near the anode to 6 μm near the edge of the device. The structure is designed to achieve a gradual reduction in the JTE charge from the anode to the edge of the device. The total length of the JTE is ~80μm.

The devices were fabricated on halide vapor phase epitaxy (HVPE) grown lightly doped ($N_D \approx 1\times10^{16}$ cm$^{-3}$) 10μm (001) epi-layer from Novel Crystal Technologies. The doping concentration of the heavily doped (001) substrate is $\approx 5\times10^{18}$ cm$^{-3}$. The fabrication of the HJDs started by evaporating Ti/Au (50/100nm) metal stack as the backside ohmic contact on the Ga$_2$O$_3$ substrate, followed by 470 °C rapid thermal annealing (RTA) for 1 minute in N$_2$. Then, a uniform 20 nm thick p-NiO layer was deposited using RF sputtering to define the active area and the first layer of the SM-JTE. NiO sputtering was carried out at an RF power of 300W and a pressure of 4 mT in pure Ar atmosphere. This was followed by a second p-NiO deposition with thickness of 40 nm to form the second layer of the SM-JTE. The NiO rings were realized by making use of a bilayer photoresist process (AZ3312/ LOR10A) followed by liftoff. The p-NiO was annealed post deposition at 300 °C for 5 minutes in N$_2$ environment. Finally, Ni/Au (50/100nm) front-contact was evaporated and annealed again at 300°C for 5 minutes in N$_2$ to form the ohmic contact to p-NiO. Charge concentration in p-NiO was measured by fabrication of C-V test structures on n$^+$ Ga$_2$O$_3$ substrate. N$_A$ extracted from C-V was around 4-5 × 10$^{18}$ cm$^{-3}$. Fig.1 (b) shows the optical image of a representative device with an active area of 75×75 μm$^2$. The undercut resulting from the use of the bi-layer photoresist resulted in the NiO layers having a small-angle bevel of ~4°, as estimated from atomic force microscopy (see Fig.1 c).

## III. Results And Discussion

The forward I-V characteristics of the diodes in linear and semi-logarithmic scales are shown in Fig. 2. (a) and (b), respectively. In Fig. 2(a), the forward current density (I$_F$) is normalized to the anode active area of 75×75 μm$^2$. The turn-on voltage (V$_{ON}$) of the HJDs estimated at I$_F$ = 1 A-cm$^{-2}$ is around 0.8V, which is significantly lower than what has been reported in literature for NiO/Ga$_2$O$_3$ HJDs [7]. The diodes also achieve a current density of 100 A-cm$^{-2}$ at a low forward bias of 1.5V, thanks to the sub-1 V V$_{ON}$. The HJDs also display a low ideality factor ($\eta$) of ~1.1, close to what is typically observed in $\beta$-Ga$_2$O$_3$ Schottky diodes [29]. However, prior reports show that NiO/Ga$_2$O$_3$ HJDs typically show a much higher ideality factor due to the trap-assisted tunneling transport mechanism [30]. Fig. 2(c) shows the comparison of forward I-V characteristics of HJDs (without field termination) with NiO deposited at O$_2$/Ar ratio of 0% and 5%. It is clear that the presence of oxygen in the sputtering gas mixture has a significant impact on the V$_{ON}$ and $\eta$ of the diodes. However, further experiments are needed to fully understand the exact mechanism behind this effect. The differential specific on-resistance (R$_{on,sp}$) of the diodes is estimated to be 4.4mΩ-cm$^{-2}$ at a current density of ~200 A/cm$^2$. As shown in Fig.2 (b), the I$_{on}$/I$_{off}$ ratio is about ~10$^{11}$ which is limited by the noise floor of the measurement set up. The charge concentration in the Ga$_2$O$_3$ epilayer is extracted using circular Schottky test structures with a diameter of 200 μm, giving N$_D$–N$_A$=1x10$^{16}$ cm$^{-3}$ as shown in

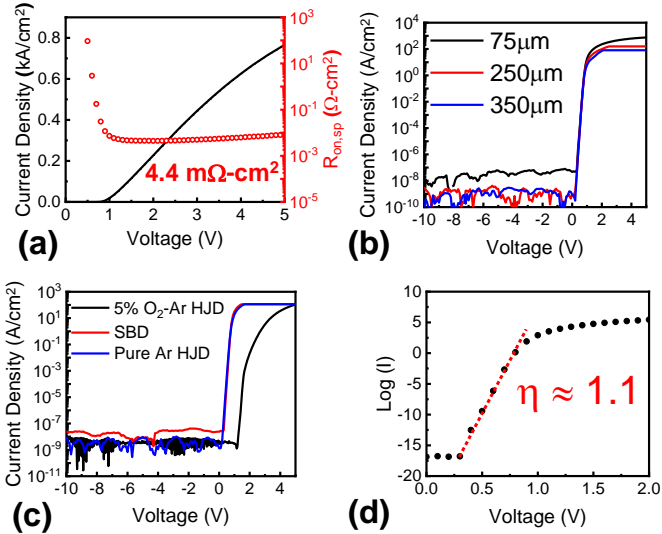

Fig. 2. (a) Forward I-V characteristics in linear scale and differential specific on-resistance for 75μm area device. (b) Forward I-V characteristics in semi-log scale for devices with 3 different active areas. (c) Forward I-V characteristics plotted in semi-log scale for SBD, pure-Ar sputtered NiO/Ga$_2$O$_3$ diode and 5%O$_2$-Ar sputtered NiO/Ga$_2$O$_3$ diodes. (d) Ideality factor extraction from semi-log I-V plot for pure-Ar sputtered NiO/Ga$_2$O$_3$ diode.

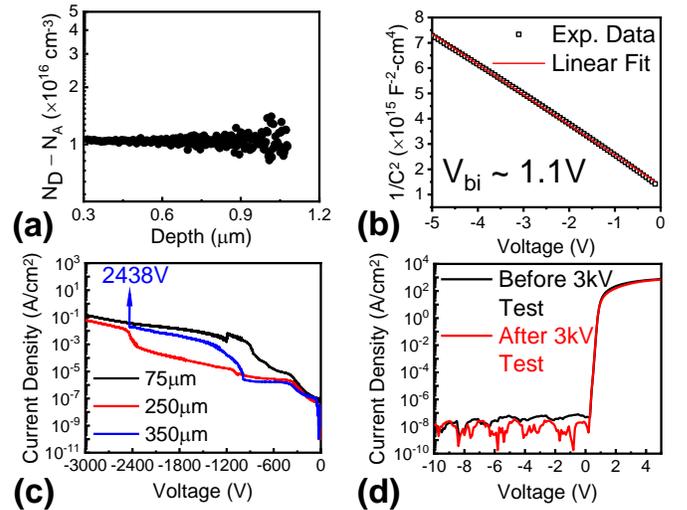

Fig. 3. (a) Plot showing charge concentration extracted from C-V measurements. (b) Built-in voltage (V$_{bi}$) extracted from C-V measurements on pure-Ar sputtered NiO/Ga$_2$O$_3$ diodes. (c) Reverse I-V characteristics plotted for devices with 3 different active areas. (d) Plot showing the comparison between 2 forward I-V curves before and after the 3kV reverse test.

Fig. 3 (a). Similar to the forward I-V, the linear extrapolation of 1/C$^2$ – V plot also results in a low built in voltage (V$_{bi}$) of 1.1 V as shown in Fig. 3 (b). The breakdown characteristics of the HJDs are shown in Fig. 3 (c). The High voltage measurements were performed using a Keysight B1505A parameter analyzer. The reverse I-V current shown in Fig. 3(c) is normalized to the total area of the device (active + JTE area). The devices with anode size of 75 µm and 250 µm displayed breakdown voltage in excess of 3 kV (measurement limit). Devices with an active area of 350µm showed a lower breakdown voltage of ~2.4 kV. As shown in Fig. 3 (d), the devices that sustained the 3 kV reverse bias measurements showed minimal change in forward I-V characteristics when measured again. Fig. 4(a) shows the electric field contour plot at a reverse bias of 3 kV obtained from Silvaco TCAD simulations for the SM-JTE diodes. Fig.

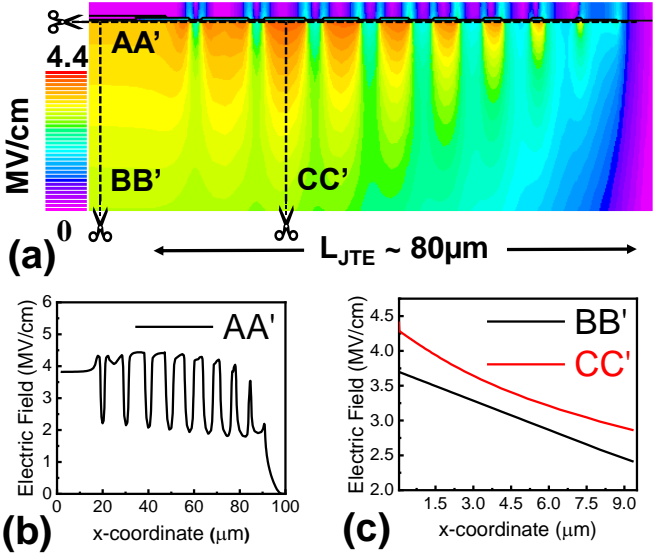

Fig. 4. (a) Electric field contour plot obtained from TCAD silvaco simulations at 3kV reverse bias. (b) Electric-field vs. x-coordinate plot for the indicated cut-line AA'. (c) Electric field vs. x-coordinate plot for the indicated cut-line BB' and CC'.

4(b) shows the electric field profile along a lateral cutline (AA') near the heterojunction interface (within the $Ga_2O_3$ epi). A nearly flat electric field profile is obtained, indicating efficient edge-termination using the SM-JTE structure. The peak electric field obtained at 3 kV is ~4.4 MV/cm, at the 3rd JTE ring. The parallel plane surface electric field at the junction at 3 kV is estimated to be 3.8 MV-cm$^{-1}$. The junction termination efficiency at 3 kV, calculated as $E_{peak}/E_{parallel-plane}$ is ~86%. Fig. 4 (c) shows the electric field profile for 2 vertical cut-lines BB' and CC' (as shown in Fig. 4(a)). The BB' and the CC' cutlines represent electric field profiles along the parallel plane and peak electric field regions The calculated Baliga's power figure-of-merit (BFOM=$V_{BR}^2/R_{on,sp}$) for our best performing devices is >2 GW-cm$^{-2}$ which is close to the theoretical unipolar FOM for SiC. We also benchmarked our diodes with vertical $Ga_2O_3$ SBDs and HJDs from the literature as shown in Fig. 5. The benchmarking was performed using both the differential $R_{on}$ (dV/dI) and the effective $R_{on}$ (V/I, at I=100 A/cm$^2$). As shown in Fig. 5 (b) we obtain an effective $R_{on}$ of 14.5mΩ-cm$^{-2}$, which is amongst the lowest reported for vertical NiO/$Ga_2O_3$ HJDs with $V_{BR}$> 1 kV.

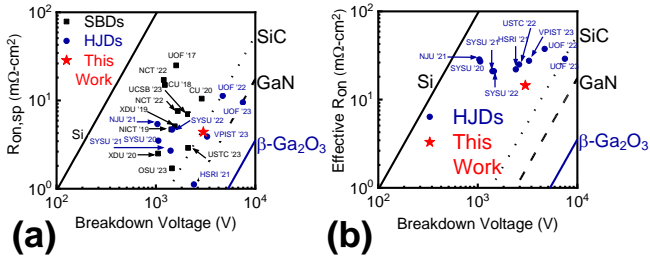

Fig. 5. (a) Benchmarking plot for the vertical SBDs, HJDs with our work with $R_{on,sp}$ vs breakdown voltage. (b) Benchmarking plot for the vertical HJDs with our work with effective $R_{on}$ vs breakdown voltage.

## IV. CONCLUSION

In conclusion, we demonstrate p-NiO/$Ga_2O_3$ HJDs with space-modulated junction termination extension (SM-JTE) enabling a gradual reduction in JTE charge. We simultaneously obtain excellent forward and reverse performance including a low $V_{ON}$ of 0.8 V, $V_{BR}$ of 3 kV and $R_{on}$ of 4.4 mΩ-cm$^{-2}$ resulting in a power figure of merit exceeding 2 GW-cm$^{-2}$. We also report a record low effective $R_{on}$ of 14.5mΩ-cm$^{-2}$ for NiO/$Ga_2O_3$ HJDs. The high-voltage reverse blocking capacity with sub-1V turn-on voltage showcases the great promise of $Ga_2O_3$ HJDs for low-loss kilovolt class power electronics.